\documentclass[amsmath,amssymb,showkeys,superscriptaddress,aps,prd,10pt,onecolumn]{revtex4-2}
\usepackage{graphicx}
\usepackage[caption=false]{subfig}
\usepackage{multirow}
\usepackage{appendix}
\usepackage{graphicx,epstopdf}
\usepackage{colortbl}
\usepackage{xcolor}
\usepackage[colorlinks=true, urlcolor=blue, linkcolor=blue, citecolor=blue]{hyperref}
\usepackage{float}
\usepackage[utf8]{inputenc}
\usepackage{graphicx}

\begin{document}

\title{Flavor Exotic Triply-Heavy Tetraquark States in AdS/QCD Potential}

\author{Halil Mutuk}
 \email{hmutuk@omu.edu.tr}
 \affiliation{Department of Physics, Faculty of Science, Ondokuz Mayis University, Atakum, 55200 Samsun, Türkiye}

\begin{abstract}
We study the $S$-wave mass spectra of flavor exotic triply-heavy tetraquark states $cc\bar{c}\bar{q}$, $cc\bar{b}\bar{q}$, $bb\bar{c}\bar{q}$ and $bb\bar{b}\bar{q}$. We adopt a diquark-antidiquark scheme to solve Schr\"{o}dinger equation. The calculations are carried out in a nonrelativistic quark model with a color interaction described by a potential computed in AdS/QCD. The AdS/QCD potential model consists of a central potential which reflects short distance and large distance behaviour of QCD, spin dependent term for hyperfine splitting and a constant term. We find stable state candidates in the $cc\bar{c}\bar{q}$ sector whereas in the $cc\bar{b}\bar{q}$, $bb\bar{c}\bar{q}$ and $bb\bar{b}\bar{q}$ sectors all the states lie above corresponding $S$-wave meson-meson thresholds.
\end{abstract}

\maketitle


\section{\label{sec:level1}Introduction}

Quark model describes ordinary mesons as ($q\bar{q}$) systems and baryons as ($qqq$) systems in terms of quark $q$ and antiquark $\bar{q}$. It explained physical properties of observed hadrons up to the millennium. In the original work of Murray Gell-Mann \cite{Gell-Mann:1964ewy}, multiquark states were conjectured. Definition of an exotic state stems from the quark model. According to the quark model, the parity for a meson is $P=(-1)^{L+1}$ and $C$ parity for a neutral meson is $C=(-1)^{L+S}$, where $L$ and $S$ are the orbital and spin angular momentum, respectively. In this picture the following combinations are allowed: $J^{PC}=0^{-+}, 0^{++}, 1^{--}, 1^{+-}, 1^{++}, \cdots$. A meson with quantum numbers that cannot be included by any combination spin-ortibal state of a quark-antiquark, such as $J^{PC}=0^{--}, 0^{+-}, 1^{-+}, 2^{+-}$, is exotic. Besides that, state(s) with unusual properties but ordinary quantum numbers are called “crypto-exotic” states \cite{Richard:2016eis}. These states can have unconventional content of valence quarks or gluons. Some hadrons do not have exotic parity, charge parity, or flavor quantum numbers but may have exotic color or flavor or spatial configurations. In addition to $J^{PC}$ exotics, flavor-exotics could also be in case.The investigation of exotic mesons which are particles either have unusual quantum numbers that are not possible in the conventional quark model or built up of four valence quarks (i.e. tetraquarks) became an important topic in the high energy physics field, especially with the observation of exotic XYZ states. The milestone was the observation of $X(3872)$ (now referred to name as $\chi_{c1}(3872)$) with a mass of ($3872.0 \pm 0.6$ MeV) in the charmonium sector \cite{Choi:2003ue}. Existence of multiquark states does not contradict to first principles of Quantum chromodynamics (QCD). The studies for exotic hadrons may open a new window for understanding of the nuclear matter and most important may provide important hints to the understanding of the nonperturbative aspects of QCD. 

After the discovery of $X(3872)$, many more charmoniumlike and bottomoniumlike states were observed. In 2020, the LHCb Collaboration observed two states, called $X_0(2900)$ and $X_1(2900)$ in the invariant mass distribution of $B^+ \to D^+ D^- K^+$ \cite{LHCb:2020pxc}. These are the first states that have four different flavors of quarks. In the same year, the observation of a charged state referred as $Z_{cs}^ -(3985)$ is anonunced by the BESIII Collaboration \cite{BESIII:2020qkh}. Very recently the first doubly-charmed tetraquark $T_{cc}^ +$ is observed by the LHCb Collaboration \cite{LHCb:2021vvq,LHCb:2021auc}. These exotic states contain at least one or two heavy quarks and triggered many phenomenological studies to enlighten their internal structures and dynamics. 

A recent observation of a narrow structure around 6.9 GeV, the so called $X(6900)$, and a broad structure around 6.2 GeV to 6.8 GeV  in the $J/\psi$ mass spectrum is a good candidate for fully-heavy tetraquarks \cite{LHCb:2020bwg}. This observation led to plenty of studies in the literature. In Ref. \cite{liu:2020eha}, this narrow structure can be explained by the $P$-wave $cc\bar{c}\bar{c}$ states in nonrelativistic quark model. In dynamical diquark model, $X(6900)$ state is assigned as $2S$ multiplet \cite{Giron:2020wpx}. Using chiral quark model and quark delocalization color screening model, $X(6900)$ is explained as a compact resonance with the quantum number $J^P=0^+$ \cite{Jin:2020jfc}.  It is argued in Ref. \cite{Zhu:2020xni} that $X(6900)$ may be interpreted as radially excited $0^{++}(3S)$ or $2^{++}(3S)$ or orbitally excited $2P$ state with quark content $cc\bar{c}\bar{c}$. Using diquark-antidiquark interpolating currents in QCD sum rule, the narrow structure is assigned to be a $P$-wave state with $J^{PC}=0^{-+}$ or $J^{PC}=1^{-+}$ and broad structure to be an $S$-wave $cc\bar{c}\bar{c}$ state with $J^{PC}=0^{++}$ or $J^{PC}=2^{++}$ \cite{Chen:2020xwe}. Ref. \cite{Yang:2020wkh} found via QCD sum rules that broad structure around 6.2-6.8 GeV can be interpreted as the $J^{PC}=0^{++}$ octet–octet tetraquark state and narrow structure around 6.9 GeV can be interpreted as $J^{PC}=0^{++}$ the octet–octet tetraquark state. The debate about the internal structure of these states is an ongoing endeavour. For more details about these structures, see a recent review  \cite{Chen:2022asf}.

If $X(6900)$ is a genuine fully-heavy tetraquark state, it is natural to think that tetraquarks with three heavy quarks $(Q)$ and one light quark $(q)$, $(QQ\bar{Q}\bar{q})$ may also form compactly. Triply-heavy tetraquarks can be obtained from fully-heavy tetraquarks by replacing one heavy quark with one light antiquark, and the strong interactions  resume to be provided by short range one-gluon exchange (OGE) and quark confinement. As mentioned in Ref. \cite{Lu:2021kut}, triply-heavy tetraquarks can be related to the doubly-heavy ones and may present a possible discrimination between singly-heavy tetraquarks and traditional heavy-light mesons. For example the inner structure of $D_{s0}^\ast(2317)$ is still undetermined. The quark content can be $c\bar{s}$, $cq\bar{s}\bar{q}$, or even $c\bar{s}-cq\bar{s}\bar{q}$ \cite{Chen:2016spr}. The problem for the nature of $D_{s0}^\ast(2317)$ arises from the fact that the light quark pair can create or annihilate dynamically in this energy region. However, the genuine tetraquarks can be easily recognized if the light quark pair is replaced with the charmonium $c\bar{c}$ or bottomonium $b\bar{b}$ pair. Traditional heavy-light mesons can become hidden charm and bottom triply-heavy tetraquarks with the excitation of one heavy-quark pair. The associated production of bottomonia and open charm hadrons $\Upsilon D/D_s$ in $pp$ collisions was reported \cite{LHCb:2015wvu}. 

There are few studies regarding triply-heavy tetraquarks in the literature compared to doubly and fully-heavy tetraquarks. Triply-heavy tetraquarks were investigated by an extended relativized quark model in Ref. \cite{Lu:2021kut} where no stable sate was reported. Ref. \cite{Silvestre-Brac:1992kaa} studied ($Q^2 \bar{Q}^2$) systems with chromagnetic interaction and reported that some configurations of triply-heavy tetraquark are possible. Ref. \cite{Silvestre-Brac:1993zem} did not find any bound triply-heavy tetraquarks using a nonrelativistic quark model. Ref. \cite{Cui:2006mp} obtained masses of possible heavy tetraquarks including ($qQ\bar{Q}\bar{Q}$) by using the color-magnetic interaction. Ref. \cite{Chen:2016ont} calculated mass splittings and masses of the $(QQ\bar{Q}\bar{q})$ tetraquark states in the framework of the color-magnetic interaction. Ref. \cite{Jiang:2017tdc} studied triply-heavy tetraquarks within the framework of QCD sum rule and $bb \bar b \bar q$ state was found to be very narrow. It was found in lattice QCD studies that the lowest spin-1 $uc\bar{b}\bar{b}$ and $sc\bar{b}\bar{b}$ states are near to their corresponding meson-meson thresholds \cite{Junnarkar:2018twb,Hudspith:2020tdf}. Nonstrange multiquark exotics as Efimov states were studied in a holographic description \cite{Liu:2019mxw}. They found that hidden-flavor tetraquark exotics such as $Q\bar Q q \bar q$, $QQ\bar Q \bar q$, and $QQ\bar Q \bar Q$ are unbound as compact topological molecules. Masses, lifetimes and weak decays of the triply heavy $b\bar{c}b\bar{q}$ tetraquarks were studied \cite{Xing:2019wil}. The mass spectrum of $S$-wave triply-heavy tetraquarks was studied by an extended chromomagnetic model \cite{Weng:2021ngd}. They found no stable state lie below the thresholds of two pseudoscalar mesons. Triply-heavy tetraquark states are investigated in the framework of the chiral quark model with the resonance ground method \cite{Liu:2022jdl}. Some stable state candidates were obtained. The results of abovementioned works are controversial: a consistent conclusion on  triply-heavy tetraquarks is still absent. 

QCD is known to be the fundamental quantum field theory of strong interactions. Charmonium $(c\bar{c})$ and bottomonium $(b\bar{b})$ states are `old boys' for the understanding of perturbative and nonperturbative natures of QCD. There are rich and precise experimental data for the traditional heavy quarkonium which help us to clarify the quark-antiquark forces \cite{Bali:2000gf,Brambilla:2010cs}. Potential models are useful tools to describe the forces acting between quarks in a phenomenologically acceptable manner \cite{Lucha:1991vn}. Short-distance behaviour of the quark-antiquark potential can be represented by a Coulomblike term while the long-distance behaviour can be represented by a linear term in the separation $r$. The quark-antiquark potential have contributed to the understanding of strong interactions and quark confinement which is beyond the reach of perturbation theory. Apart from conventional mesons and baryons, QCD foresees multiquark states which may provide important hints for discerning of the nonperturbative aspects of QCD. There are many studies for tetraquark states in the literature using potential models, see for the recents \cite{Yang:2020rih, Zhao:2021cdv, Asadi:2021ids,Yang:2021zhe}.

The AdS/CFT correspondence relates nonperturbative aspects of QCD with string theory in order to get an analytical first order approach to nonperturbative effects. It also provides insights into the nonperturbative part of QCD such as the orbital and radial spectra of hadrons \cite{Brodsky:2008pg}. In this respect AdS/QCD models find place in the literature. In Ref. \cite{Kim:2007rt}, charmonium vector states were studied in hard-wall, soft-wall and brainless setup of AdS/QCD models at zero and finite temperatures. Using the static potential inspired by holographic principle with both the vanila AdS-Schwarzschild metric and the one with an infrared cutoff, dissociation temperatures of heavy quarkonia were calculated \cite{Hou:2007uk}. In Ref. \cite{Bai:2013rza}, a holographic model of heavy-light mesons by extending the AdS/QCD correspondence to incorporate the behavior of the heavy quark limit was constructed.  Baryon spectrum at finite temperature by using AdS/QCD correspondence is calcualted in Ref. \cite{Li:2013lfa}. Ref. \cite{Wang:2015osq} studied meson, nucleon and $\Delta$-baryon properties at zero and finite temperatures in the hard-wall AdS/QCD model. Phenomenological analysis of the scalar glueball and scalar meson spectra within an AdS/QCD framework  is presented in Ref. \cite{Rinaldi:2020ssz}. $S$-wave charmonium and bottomonium states were described at zero and finite temperatures by using bottom-up AdS/QCD model \cite{MartinContreras:2021bis}. 

In this present paper, we use AdS/QCD inspired potential in diquark-antidiquark approach to study spectroscopy of triply-heavy tetraquark states. This potential is derived in the context of AdS/QCD correspondence and have similar behavior as expected from QCD \cite{Andreev:2006ct}. It was successfully used to study pentaquark states \cite{Giannuzzi:2019esi}, meson and tetraquark states \cite{Carlucci:2007um}, radiative decays of $\eta_c$ and $\eta_b$ mesons \cite{Giannuzzi:2008pv}, and doubly heavy baryons \cite{Giannuzzi:2009gh}. In a recent study, the energy configuration in a quark-antiquark pair from Nambu-Goto action in a deformed AdS space is calculated \cite{Bruni:2018dqm}. It was shown that this configuration has the shape of a Cornell potential. To the best of our knowledge, there is no reported work on triply-heavy tetraquark states using AdS/QCD potential.

This paper is organized as follows: In Sections \ref{sec:level2} and \ref{sec:level3}, we describe AdS/QCD potential model and diquark-antidiquark model of this work, respectively. In Section \ref{sec:level4}, the masses of the triply-heavy tetraquark systems are calculated and comparison of triply-heavy tetraquark masses with previous studies within different approaches are given. Section \ref{sec:level5} is reserved for conclusion and summary of the obtained results. 

\section{\label{sec:level2}AdS/QCD Potential}

Heavy quarkonium bound states are important systems for our understanding of the QCD due to the their nonrelativistic nature. In this perspective, it is possible to describe the interaction between quark $q$ and  antiquark $\bar{q}$ in terms of a local potential $V(r)$, where $r$ is the interquark separation. The mostly used potential in literature is the so called Cornell potential \cite{Eichten:1978tg}. There are also other potential models in the literature which gave successful results \cite{Richardson:1978bt,Buchmuller:1980su,Bhaduri:1981pn,Semay:1994ht}. These potentials may include some spin dependent and constant terms in addition to Coulombic and linear parts. 

Using AdS/QCD correspondence, the $(Q\bar{Q})$ potential was derived in \cite{Andreev:2006ct}. This potential was used to obtain meson spectra as mentioned before. The AdS/QCD correspondence is a new approach for studying the nonperturbative regime of QCD. We shall use this potential with the corresponding parameters of Ref. \cite{Carlucci:2007um}. The potential can be written as
\begin{equation}
V(r) = V_{AdS/QCD}(\lambda)+V_{spin}(r)+V_0,
\label{eqn:TotPot}
\end{equation}
where $V_{AdS/QCD}$ is the AdS/QCD potential, $V_{spin}$ is the spin-spin interaction term and $V_0$ is a constant. The $V_{AdS/QCD}$ is given as a parametric equation

\begin{eqnarray}
\label{eqn:AdSPot}
V_{AdS/QCD}(\lambda)&=&\frac{g}{\pi}\sqrt{\frac{c}{\lambda}} \left( -1+ \int_0^1 dv v^{-2}  \left[ e^{\lambda v^2/2} \left( 1-v^4 e^{\lambda(1-v^2)} \right)^{-1/2} -1\right] \right), \nonumber \\
r(\lambda)&=& 2\sqrt{\frac{\lambda}{c}}\int_0^1 dv v^{2} e^{\lambda v^2/2} \left( 1-v^4 e^{\lambda(1-v^2)} \right)^{-1/2},
\end{eqnarray}
where $r$ is the interquark distance and $\lambda$ is in the range of $0\leq \lambda < 2$. This interquark potential derivation is conducted in the framework of the gauge/string duality approach \cite{Maldacena:1997re,Maldacena:1998im,Rey:1998ik}. The $V_{spin}(r)$ spin term is responsible for the spin-spin interactions
\begin{equation}
V_{spin}(r)=A\frac{\delta(r)}{m_1 m_2} (\textbf{S}_1 \cdot \textbf{S}_2) ~ \text{with}  ~ \delta(r)= \left( \frac{\sigma}{\sqrt{\pi}} \right)^3 e^{-\sigma^2r^2},
\end{equation}
where $\sigma$ is a parameter of the smeared delta function and $A$ is the parameter proportional to the QCD running coupling constant and gets two different values: $A_b$ in the case of mesons comprising at least bottom quark and $A_c$ for otherwise. 

In the center of mass frame, it is possible to factorize the angular and radial parts of the Schr\"{o}dinger  equation as
\begin{equation}
\bigg\{ \frac{1}{2\mu} \left[-\frac{d^2}{dr^ 2} + \frac{L(L+1)}{r^2} \right] +V(r) \bigg\} \psi(r)=E \psi(r).
\end{equation}
Here $\mu=m_1m_2/(m_1+m_2)$ is the reduced masses of quark 1 and quark 2. In Refs. \cite{Giannuzzi:2019esi,Carlucci:2007um,Giannuzzi:2008pv,Giannuzzi:2009gh} a cutoff is introduced in the context of Bethe-Salpeter equation to cure the singularity since as $\lambda \to 0$ the potential diverges in Eq. \ref{eqn:AdSPot}. This divergence is harmless in the context of Schr\"{o}dinger equation \cite{Carlucci:2007um}.  We solve Schr\"{o}dinger equation numerically by assuming Dirichlet boundary conditions. 

\section{\label{sec:level3}Diquark Model}

The history of diquark is old as quarks. Diquark structure was first suggested by Gell-Mann \cite{Gell-Mann:1964ewy}. A diquark is a bound quark-quark $(qq)$ pair, whereas an antidiquark is a bound antiquark-antiquark $(\bar{q}\bar{q})$ pair. This binding is a result of OGE between the quarks. 

The diquark structure is an important object for the underlying structure of baryons. Diquark model for baryon was introduced in Refs. \cite{Ida:1966ev,Lichtenberg:1967zz,Anselmino:1992vg} such that a diquark can form inside the baryon. This quark-diquark approach was successfully applied for baryons, see for example \cite{Santopinto:2004hw,Ebert:2007nw,Faustov:2015eba,MoosaviNejad:2020nsl,Faustov:2021qqf}. Diquarks are also important for the understanding of tetraquarks and mesons outside of scope of the quark model. Many models in the literature explain the inner structure of tetraquarks to be compact and to consist of diquark-antidiquark pairs \cite{Ali:2019roi}. In this perspective it can be said that tetraquarks are composed of diquarks and antidiquarks. These diquarks and antidiquarks are not by themselves colorless but are introduced as constituent objects of tetraquark systems to form colorless combinations. Indeed, there are some indications of diquark correlations in the hadrons (for a recent review see \cite{Barabanov:2020jvn}). 

In the study of triply-heavy tetraquarks, we will treat the four-body problem as three two-body problem. Considering the diquarks and antidiquarks as constituents of the tetraquarks to predict the tetraquark masses were done in many works, see for example Refs. \cite{Maiani:2004vq,Ebert:2005nc,Lundhammar:2020xvw,Monemzadeh:2014vhh,Lu:2016zhe,Debastiani:2017msn,Bedolla:2019zwg,Ghalenovi:2020zen,Faustov:2020qfm,Giron:2020wpx}. At first step we compute the mass of the diquark, then we do the same for antidiquark and finally we solve the Schr\"{o}dinger equation once again for a two-body system composed of a diquark and an antidiquark. For the single-flavor heavy tetraquarks, the second step is simultaneously done when the first step is over, i.e. the mass of the diquark equals to the mass of the antidiquark. But in the triply-heavy tetraquarks, the mass of the diquark is not equal to the antidiquark due to the quark content.  

Diquark masses are needed as input to compute triply-tetraquark masses. When we combine two quarks in the fundamental color representation, it reduces to $\vert q q \rangle:  \text{3}  \bigotimes  \text{3}= \bar{\text{3}} \bigoplus \text{6} $, a color antitriplet  $\bar{\text{3}}$ and a color sextet $\text{6}$. In a similar way, combining two antiquarks reduces to $\vert \bar{q} \bar{q} \rangle:  \bar{\text{3}}  \bigotimes  \bar{\text{3}}= \text{3} \bigoplus \bar{\text{6}} $, a triplet $\text{3}$ and antisextet $\bar{\text{6}}$. The antitriplet state is attractive whereas the sextet state is repulsive. Combining an antitriplet diquark and a triplet antidiquark form a color singlet tetraquark, $3-\bar{3}$. In addition to this configuration, $6-\bar{6}$ configuration can also yield a tetraquark state. 

In OGE approximation, the attractive potential between two quarks in diquark is half the one between a quark and antiquark. Therefore, it is natural to use the following relation for the potential of diquark
\begin{equation}
V_{QQ}=\frac{1}{2} V_{Q\bar{Q}}.
\end{equation}
Many authors also extended this factor 1/2 to the whole potential describing the quark-quark interaction in different models  \cite{Ebert:2007rn,Lu:2016zhe,Debastiani:2017msn}. In the sextet configuration, the form of the diquark potential derived from the quark-antiquark potential changes. This will be explained in the following section.

In this paper, we consider a diquark with internal orbital quantum number $\ell =0$. We also assume that diquarks are pointlike particles. The mass spectrum for triply-heavy tetraquarks is obtained using
\begin{equation}
M=m_d + m_{\bar{d}} + E,
\end{equation}
where $m_d$ is the mass of diquark, $m_{\bar{d}}$ is the mass of antidiquark, and $E$ is the energy eigenvalue. 

Constructing $\text{color} \otimes \text{spin} $ wave functions for the triply-heavy tetraquark states can be done under the Pauli exclusion principle as
\begin{eqnarray}
\phi_1 \chi_1 &=& \vert (Q_1 Q_2)_0^{\bar{3}}(\bar{Q}_3\bar{q}_4)^{3}_0\rangle_0\delta_{12}, \nonumber \\ 
\phi_1 \chi_2 &=& \vert (Q_1Q_2)_1^{\bar{3}}(\bar{Q}_3\bar{q}_4)^{3}_1\rangle_0,\nonumber \\
\phi_1 \chi_3 &=& \vert (Q_1Q_2)_0^{\bar{3}}(\bar{Q}_3\bar{q}_4)^{3}_1\rangle_1\delta_{12},\nonumber \\
\phi_1 \chi_4 &=& \vert (Q_1Q_2)_1^{\bar{3}}(\bar{Q}_3\bar{q}_4)^{3}_0\rangle_1,\nonumber \\
\phi_1 \chi_5 &=& \vert (Q_1Q_2)_1^{\bar{3}}(\bar{Q}_3\bar{q}_4)^{3}_1\rangle_1,\nonumber \\
\phi_1 \chi_6 &=& \vert (Q_1Q_2)_1^{\bar{3}}(\bar{Q}_3\bar{q}_4)^{3}_1\rangle_2,\nonumber \\
\phi_2 \chi_1 &=& \vert (Q_1Q_2)_0^6(\bar{Q}_3\bar{q}_4)_0^{\bar{6}}\rangle_0,\\ \label{colorspin}
\phi_2 \chi_2 &=& \vert (Q_1Q_2)_1^6(\bar{Q}_3\bar{q}_4)_1^{\bar{6}}\rangle_0\delta_{12}, \nonumber \\
\phi_2 \chi_3 &=& \vert (Q_1Q_2)_0^6(\bar{Q}_3\bar{q}_4)_1^{\bar{6}}\rangle_1, \nonumber \\
\phi_2 \chi_4 &=& \vert (Q_1Q_2)_1^6(\bar{Q}_3\bar{q}_4)_0^{\bar{6}}\rangle_1\delta_{12},\nonumber \\
\phi_2 \chi_5 &=& \vert (Q_1Q_2)_1^6(\bar{Q}_3\bar{q}_4)_1^{\bar{6}}\rangle_1\delta_{12}, \nonumber\\
\phi_2 \chi_6 &=& \vert (Q_1Q_2)_1^6(\bar{Q}_3\bar{q}_4)_1^{\bar{6}}\rangle_2\delta_{12},\nonumber 
\end{eqnarray}
where $\delta_{12}=0$ if $Q_1$ and $Q_2$ are identical flavor quarks, and $\delta_{12}=1$ for the other cases. Here superscript 3, $\bar{3}$, 6 or $\bar{6}$ denotes the color, and the subscript 0, 1 or 2 denotes the spin. For a good discussion of the color and spin wave functions of the heavy tetraquarks, see Ref. \cite{Park:2013fda}. 

To exhaust all possible configurations of the triply-heavy tetraquarks $(QQ\bar{Q}\bar{q})$, it is legitimate to replace heavy quark $Q$ by either charm $c$ or bottom $b$ quark and light quark $q$ by $u$, $d$ and $s$. The possible cases we have are $bb\bar{b}\bar{q}$, $bb\bar{c}\bar{q}$, $bc\bar{b}\bar{q}$, $bc\bar{c}\bar{q}$, $cc\bar{c}\bar{q}$, and $cc\bar{b}\bar{q}$. In this paper we study flavor-exotic $cc\bar{c}\bar{q}$, $cc\bar{b}\bar{q}$, $bb\bar{c}\bar{q}$ and $bb\bar{b}\bar{q}$ states. The $cb\bar{c}\bar{q}$ and $cb\bar{b}\bar{q}$ systems are quite different systems. $cb\bar{c}\bar{q}$ and $cb\bar{b}\bar{q}$ tetraquarks are hidden charm and hidden bottom tetraquarks, respectively and no flavor exotic state exists in these configurations.

\section{\label{sec:level4}Mass Spectrum}

The parameters of the model are as follows: $c=0.3 ~ \text{GeV}^2$, $g=2.75$, $V_0=-0.49 ~ \text{GeV}$, $A_c=7.92 ~ \text{GeV}^3$, $A_b=3.09 ~ \text{GeV}^3$, $k=1.48$, $k^\prime=2.15$, $\sigma=1.21~ \text{GeV}^2$, $m_q=0.302 ~\text{GeV}$, $m_s=0.454 ~\text{GeV}$, $m_c=1.733 ~\text{GeV}$, $m_b=5.139 ~\text{GeV}$ \cite{Carlucci:2007um}. The potential in Eq. \ref{eqn:TotPot} can be used in a quark model to calculate tetraquark spectra under the hypothesis that a tetraquark can be considered as a bound state of a diquark and an antidiquark. 

\subsection{Diquark Masses}
Before calculating masses of tetraquark states, we will first obtain diquark masses. The results are listed in Table \ref{diquarkmass}. As mentioned before, the parameter $A$ is related to the strong coupling constant $\alpha_s$, in our case $A \sim \kappa \alpha_s$, and string tension is related to $c$. For triplet configurations of diquarks, the potential is half of the quark-antiquark potential. For sextet configurations, color factor and string tension change as in the triplet configuration. It is given in Ref. \cite{Debastiani:2017msn} that color factor $\kappa$ of Coulomb term in the Cornell potential, $V(r)=\frac{\kappa \alpha_s}{r} + br$, corresponding to the sextet configurations is +1/3. For the string tension $b$ of linear term in the Cornell potential, it is divided by four and sign is also changed. For more discussion see the mentioned reference. In this work, we rely on these observations and make the relevant changes in the potential for sextet configurations.  Having calculated the masses for the diquarks and antidiquarks, one can use the initial stage of the model which describe quark-antiquark systems, to study tetraquark systems.

\begin{table}[h]
\caption{\label{diquarkmass}Results for the diquark masses. The braces $\{ \}$ show symmetric wave functions of the subsystems and the parantheses () show without permutation symmetries. All results are in MeV.}
\begin{ruledtabular}
\begin{tabular}{cccc}
Configuration&Mass&Configuration&Mass\\
\hline

$(\bar c \bar n)^3_1$ & 2021& $(\bar b \bar n)^3_1$& 5342\\
$(\bar c \bar n)^3_0$ & 1982& $(\bar b \bar n)^3_0$&5325\\
$(\bar c \bar n)^{\bar 6}_1$ & 1834 & $(\bar b \bar n)^{\bar 6}_1$ &5095 \\
$(\bar c \bar n)^{\bar 6}_0$ & 1836 & $(\bar b \bar n)^{\bar 6}_0$ &5098\\
$(\bar c \bar s)^3_1$ & 2103 & $(\bar b \bar s)^3_1$ &5414\\
$(\bar c \bar s)^3_0$ & 2073 & $(\bar b \bar s)^3_0$& 5403\\
$(\bar c \bar s)^{\bar 6}_1$ & 1900& $(\bar b \bar s)^{\bar 6}_1$& 5300 \\
$(\bar c \bar s)^{\bar 6}_0$ & 1901& $(\bar b \bar s)^{\bar 6}_0$ &5302 \\
$\{c c\}^{\bar 3}_1$ & 3130 & $\{ bb\}^{\bar 3}_1$ &9646\\
$\{cc\}^6_0$ & 2917 & $\{bb\}^6_0$ &9590\\
\end{tabular}
\end{ruledtabular}
\end{table}

\subsection{The $cc\bar{c}\bar{q}$ and $cc\bar{b}\bar{q}$ Tetraquark States}

The mass spectra and possible $S$-wave thresholds are listed in Table \ref{Table:cccq}. Here we assume $SU(2)$ flavor symmetry exist for up and down quarks as $n={u,d}$ and  $q=s$. We also show values of the difference of the triply-heavy tetraquark $(M)$ and threshold masses $(T)$, $\Delta=M-T$. If this quantity is negative, then the triply-heavy tetraquark state lies below the threshold of the fall-apart decay into two-meson and thus should be a narrow state. The states with small positive values of $\Delta$ could be also observed as resonances, since their decay rates will be suppressed by the phase space. All other states are expected to be broad and thus difficult to observe in the experiments.

\begin{table}[h!]
\begin{center}
\caption{\label{Table:cccq} Predicted mass spectra for the $cc\bar c \bar q$ and $cc\bar b \bar q$ systems, corresponding thresholds and $\Delta$ values. The meson masses are taken from \cite{ParticleDataGroup:2020ssz} and the predicted mass for $M_{B_c^\ast}=6338$ MeV is taken from \cite{Godfrey:1985xj}. All results are in units of MeV.}
\begin{tabular*}{18cm}{@{\extracolsep{\fill}}p{1cm}<{\centering}p{1.7cm}<{\centering}p{4.5cm}<{\centering}p{1.8cm}<{\centering}p{4.5cm}<{\centering}}
\hline\hline
 $J^{P}$  & Configuration                                             & Mass &  $S$-wave threshold & $\Delta$ \\\hline

 $0^{+}$  & $|\{cc\}^{\bar 3}_1 (\bar c \bar n)^3_1 \rangle_0$    & \multirow{1}{*}{$\begin{bmatrix} 5005\\ 5154 \end{bmatrix}$}
               & $\begin{bmatrix} \eta_c D, J/\psi D^\ast \end{bmatrix}$ & \multirow{2}{*}{$\begin{bmatrix}(156, -99)\\(305, 50) \end{bmatrix}$}\\
            &  $|\{cc\}^6_0 (\bar c \bar n)^{\bar 6}_0\rangle_0$    \\
 $1^{+}$  &  $ |\{cc\}^{\bar 3}_1 (\bar c \bar n)^3_0\rangle_1  $    & \multirow{3}{*}{$\begin{bmatrix}5053 \\5081 \\ 5106  \end{bmatrix}$}
               & $\begin{bmatrix} \eta_c D^\ast, J/\psi D^\ast \end{bmatrix}$  & \multirow{3}{*}{$\begin{bmatrix}(62, -51)\\(90, -23) \\   (115, 2) \end{bmatrix}$}\\
                 &  $|\{cc\}^{\bar 3}_1 (\bar c \bar n)^3_1\rangle_1$    \\
            &  $|\{cc\}^{6}_0 (\bar c \bar n)^{\bar 6}_1\rangle_1$    \\
 $2^{+}$  &   $|\{cc\}^{\bar 3}_1 (\bar c \bar n)^3_1\rangle_2$    &\multirow{1}{*} {$\begin{bmatrix} 5116 \end{bmatrix}$}    & $\begin{bmatrix}J/\psi D^\ast \end{bmatrix}$  & \multirow{1}{*} {$\begin{bmatrix}  12 \end{bmatrix}$} \\\hline

 $0^{+}$  & $|\{cc\}^{\bar 3}_1 (\bar c \bar s)^3_1 \rangle_0$    & \multirow{2}{*}{$\begin{bmatrix}5112 \\ 5249 \end{bmatrix}$}
               & $\begin{bmatrix} \eta_c D_s, J/\psi D^\ast_s \end{bmatrix}$ & \multirow{2}{*}{$\begin{bmatrix}(160, -97)\\(297, 40) \end{bmatrix}$}\\
            &  $|\{cc\}^6_0 (\bar c \bar s)^{\bar 6}_0\rangle_0$    \\
 $1^{+}$  &  $ |\{cc\}^{\bar 3}_1 (\bar c \bar s)^3_0\rangle_1  $    & \multirow{3}{*}{$\begin{bmatrix}5118 \\5173 \\ 5190  \end{bmatrix}$}
               & $\begin{bmatrix} \eta_c D_s^\ast, J/\psi D^\ast_s \end{bmatrix}$  & \multirow{3}{*}{$\begin{bmatrix}(22, -91)\\(77, -36) \\   (94, -19) \end{bmatrix}$}\\
                 &  $|\{cc\}^{\bar 3}_1 (\bar c \bar s)^3_1\rangle_1$    \\
            &  $|\{cc\}^{6}_0 (\bar c \bar s)^{\bar 6}_1\rangle_1$    \\
 $2^{+}$  &   $|\{cc\}^{\bar 3}_1 (\bar c \bar s)^3_1\rangle_2$    & \multirow{1}{*} {$\begin{bmatrix}  5202 \end{bmatrix}$}  &  $\begin{bmatrix} J/\psi D^\ast_s \end{bmatrix}$ &  \multirow{1}{*} {$\begin{bmatrix}  -7 \end{bmatrix}$} \\\hline \hline

 $0^{+}$  & $|\{cc\}^{\bar 3}_1 (\bar b \bar n)^3_1 \rangle_0$    & \multirow{2}{*}{$\begin{bmatrix}8526  \\ 8638 \end{bmatrix}$}
               & $\begin{bmatrix} B_c D, B_c^\ast D^\ast \end{bmatrix}$ & \multirow{2}{*}{$\begin{bmatrix}(382, 181)\\(494, 293)\end{bmatrix}$}\\
            &  $|\{cc\}^6_0 (\bar b \bar n)^{\bar 6}_0\rangle_0$    \\
 $1^{+}$  &  $ |\{cc\}^{\bar 3}_1 (\bar b \bar n)^3_0\rangle_1  $    & \multirow{3}{*}{$\begin{bmatrix}8541 \\8598 \\ 8613  \end{bmatrix}$}
               & $\begin{bmatrix} B_c D^\ast , B_c^\ast D^\ast \end{bmatrix} $  & \multirow{3}{*}{$\begin{bmatrix}(310, 196)\\(367, 253) \\   (382, 268) \end{bmatrix}$}\\
                 &  $|\{cc\}^{\bar 3}_1 (\bar b \bar n)^3_1\rangle_1$    \\
            &  $|\{cc\}^{6}_0 (\bar b \bar n)^{\bar 6}_1\rangle_1$    \\
 $2^{+}$  &   $|\{cc\}^{\bar 3}_1 (\bar b \bar n)^3_1\rangle_2$    &  \multirow{1}{*} {$\begin{bmatrix}  8630 \end{bmatrix}$}  & $\begin{bmatrix}B_c^\ast D^\ast \end{bmatrix}$   &  \multirow{1}{*} {$\begin{bmatrix}  285 \end{bmatrix}$} \\\hline 

 $0^{+}$  & $|\{cc\}^{\bar 3}_1 (\bar b \bar s)^3_1 \rangle_0$    & \multirow{2}{*}{$\begin{bmatrix} 8576 \\ 8670 \end{bmatrix}$}
               & $\begin{bmatrix} B_c D_s, B_c^\ast D_s^\ast \end{bmatrix}$  & \multirow{2}{*}{$\begin{bmatrix}(333, 126)\\(427, 220) \end{bmatrix}$}\\
            &  $|\{cc\}^6_0 (\bar b \bar s)^{\bar 6}_0\rangle_0$    \\
 $1^{+}$  &  $ |\{cc\}^{\bar 3}_1 (\bar b \bar s)^3_0\rangle_1  $    & \multirow{3}{*}{$\begin{bmatrix}8610\\8644\\ 8651  \end{bmatrix}$}
               & $\begin{bmatrix} B_c D_s^\ast, B_c^\ast D_s^\ast \end{bmatrix}$ & \multirow{3}{*}{$\begin{bmatrix}(223, 160)\\(257, 194) \\   (264, 201) \end{bmatrix}$}\\
                 &  $|\{cc\}^{\bar 3}_1 (\bar b \bar s)^3_1\rangle_1$    \\
            &  $|\{cc\}^{6}_0 (\bar b \bar s)^{\bar 6}_1\rangle_1$    \\
 $2^{+}$  &   $|\{cc\}^{\bar 3}_1 (\bar b \bar s)^3_1\rangle_2$    & \multirow{1}{*} {$\begin{bmatrix} 8674 \end{bmatrix}$}  &  $ \begin{bmatrix}B_c^\ast D_s^\ast \end{bmatrix}$ & \multirow{1}{*} {$\begin{bmatrix}  224 \end{bmatrix}$}  \\

\hline\hline
\end{tabular*}
\end{center}
\end{table}

As can be seen from Table \ref{Table:cccq}, the masses of the hidden charm $cc\bar{c}\bar{n}$ states lie in the range of 5005-5154 MeV and $cc\bar{c}\bar{s}$ states lie in the range of 5112-5249 MeV. The mass splitting between the $cc\bar{c}\bar{n}$ tetraquarks is 149 MeV whereas it is 137 MeV in the  $cc\bar{c}\bar{s}$ tetraquarks. The mass splittings for these triply-heavy tetraquarks arises from spin-spin interactions, as in the case of conventional mesons. For the $cc\bar{b}\bar{n}$ system, the masses lie in the range of 8526-8638 MeV whereas for $cc\bar{b}\bar{s}$ system the masses lie in the range of 8576-8674 MeV. The mass splittings between $cc\bar{b}\bar{n}$ and $cc\bar{b}\bar{s}$ tetraquarks are 112 and 98 MeV, respectively. As mentioned in Ref. \cite{Chen:2016ont}, the quantum numbers of $cc\bar{c}\bar{n}$ and $cc\bar{c}\bar{s}$ tetraquark states are the same of $D$ and $D_s$ mesons. However the masses of $D$ and $D_s$ mesons, which are around 2.0-2.7 GeV, are much lower than the predicted mass values of $cc\bar{c}\bar{n}$  and $cc\bar{c}\bar{s}$ states. No orbital or radial excitation can induce such a mass difference. Therefore once these states could be observed, it is easy to identify them as $D-$ or $D_s-$like meson with an excited $c\bar{c}$ pair. It will be good to touch on the color configurations of triply-heavy tetraquark states. In Refs. \cite{Chen:2016ont,Weng:2021ngd} no explicit color configurations are presented however it was mentioned that ground states of $cc\bar{c}\bar{q}$ and $cc\bar{b}\bar{q}$ are both dominated by color-sextet $6 \otimes \bar{6}$ color configurations. For example in Ref. \cite{Weng:2021ngd}, it was pointed out that the $nc\bar{c}\bar{c}$ state with $J^P=0^+$ has a mass of $4936.7$ MeV with $66.7\%$ of the $6 \otimes \bar{6}$ component, and the state $sc\bar{c}\bar{c}$ with $J^P=0^+$ has a mass of $5040.1$ MeV with $66.6\%$ $6 \otimes \bar{6}$ component. In Ref. \cite{Lu:2021kut}, $6 \otimes \bar{6}$ color configurations have higher mass values of both $J^P=0^ +$ and $J^P=1^ +$ for $cc\bar{c}\bar{q}$ and $cc\bar{b}\bar{q}$ tetraquarks. Our results also support this conclusion except $cc \bar b \bar s$ case where $\bar{3} \otimes 3 $ with $J^P=2^+$ state is just 4 MeV above than the $6 \otimes \bar{6}$  state with $J^P=0^+$ pattern. 

Comparing to the thresholds of $cc\bar{c}\bar{q}$ and $cc\bar{b}\bar{q}$ systems we find that there are some bound state candidates in the $cc\bar{c}\bar{q}$ sector. There are some configurations in the $cc\bar{c}\bar{q}$ sector which are just few MeVs above than the corresponding two-meson thresholds. There may be a chance for observing them as resonances. The predicted masses of $cc\bar{b}\bar{q}$ states are much higher than their corresponding two-meson thresholds which refer that they may easily decay into two mesons via the fall-apart mechanism. 

We compare our mass results with the predictions of other studies in Table \ref{Table:3}. Before going into discussion, it will be good to mention that we present mass results starting from lower values for each spin state. It can be observed that for the $cc\bar{c}\bar{q}$ tetraquark states our results agree well with the results of Ref. \cite{Weng:2021ngd} in which an extended chromomagnetic model is used and with the results of Ref. \cite{Jiang:2017tdc} which is a QCD sum rule study. In Ref. \cite{Jiang:2017tdc} for the $J^P=0^+$ and $J^P=1^+$, both $6 \otimes \bar{6}$ and $\bar{3} \otimes 3$ type interpolating currents gave mass value as $5.1 \pm 0.2$ GeV for the $cc\bar{c}\bar{q}$ tetraquark states. Our results are quite lower than the predicted masses of Refs. \cite{Chen:2016ont,Lu:2021kut} for $cc\bar{c}\bar{q}$ tetraquarks. Ref. \cite{Lu:2021kut} used an extended relativistic quark model whereas in Ref. \cite{Chen:2016ont} color-magnetic interaction model is used. In the $cc\bar{b}\bar{q}$ tetraquark states, our results are compatible with the results of Ref. \cite{Lu:2021kut}. Results of Ref. \cite{Chen:2016ont} are approximately 200 MeV below than our results whereas results of Ref. \cite{Weng:2021ngd} are approximately 300 MeV above than our results. Extracted central mass values of Ref. \cite{Jiang:2017tdc} are lower than our mass values. For $J^P=0^+$, the extracted masses are $8.0 \pm 0.3$ GeV for $6 \otimes \bar{6}$ current and $8.2 \pm 0.3$ GeV for $\bar{3} \otimes 3$  current whereas for $J^P=1^+$, $6 \otimes \bar{6}$ type current gave mass as $8.1 \pm 0.3$ GeV and $\bar{3} \otimes 3$ type current gave mass as $8.2 \pm 0.3$ GeV.

\begin{table}[h!]
\caption{\label{Table:3} Comparison of $cc\bar{c}\bar{q}$ and $cc\bar{b}\bar{q}$ tetraquark masses with other studies.  All results are in units of MeV except Ref. \cite{Jiang:2017tdc} where they are in GeV.  }
\begin{ruledtabular}
\begin{tabular}{ccccccccc}
Configuration& $J^P$&This work&\cite{Chen:2016ont}&\cite{Jiang:2017tdc} &\cite{Lu:2021kut}&\cite{Weng:2021ngd} \\
\hline
 $cc\bar{c}\bar{n}$&  &  &  &  \\
 \hline
 &$0^+$&5005-5154&5411.6-5658.8& $5.1 \pm 0.2$-$5.1 \pm 0.2$ & 5403-5533 & 4936.7-5185.3\\

 &$1^+$&5053-5081-5106&5463.4-5530.5-5605.6& $5.1 \pm 0.2$-$5.1 \pm 0.2$ & 5400-5473-5482 & 4968.1-5135.5-5154.4\\

 &$2^+$&5116&5593 & $\cdots$ & 5506 & 5198.4\\ \hline

$cc\bar{c}\bar{s}$&  &  &  &  \\ \hline

 &$0^+$&5112-5249&5592.1-5839.5& $\cdots$ & 5476-5599 & 5040.1-5290.6\\

 &$1^+$&5118-5173-5190& 5638.8-5710.5-5781.9& $\cdots$ & 5481-5552-5554 & 5069.2-5239.9-5254.3\\

 &$2^+$&5202&5773.2 & $\cdots$ & 5584 & 5303.3\\ \hline \hline

 $cc\bar{b}\bar{n}$&  &  &  &  \\ \hline

 &$0^+$&8526-8638&8756.8-8962.4& $8.0\pm0.3-8.2\pm0.3$ & 8658-8761 & 8199.1-8444.8\\

 &$1^+$&8541-8598-8613& 8803.7-8868.3-8935.0& $8.1\pm0.3-8.2\pm0.3$ & 8667-8734-8740& 8217.5-8430.0-8454.4\\

 &$2^+$&8630&8908.6& $\cdots$ & 8755 & 8477.9\\ \hline 
 
 $cc\bar{b}\bar{s}$&  &  &  &  \\ \hline

 &$0^+$&8576-8670&8935.1-9140.7& $\cdots$ & 8751-8845 & 8300.5-8538.7\\

 &$1^+$&8610-8644-8651& 8982.6-9046.9-9114.0& $\cdots$ & 8761-8820-8825& 8324.8-8521.9-8553.3\\

 &$2^+$&8674&9086.9& $\cdots$ & 8840 & 8569.1\\ 

\end{tabular}
\end{ruledtabular}
\end{table}

\subsection{The $bb\bar{c}\bar{q}$ and $bb\bar{b}\bar{q}$ Tetraquark States}

The masses of $bb\bar{c}\bar{q}$ and $bb\bar{b}\bar{q}$ tetraquarks with corresponding thresholds and $\Delta$ values are presented in Table \ref{Table:bbcbq}. The predicted masses of $bb \bar c \bar q$ tetraquarks lie around 12 GeV, specifically in the range of 11949-12088 MeV for $bb \bar c \bar n$ tetraquark states and in the range of 12022-12143 MeV for $bb \bar c \bar s$ tetraquark states. The mass splittings are 139 and 121 MeV for the $bb \bar c \bar n$ and $bb\bar{c}\bar{s}$ tetraquarks, respectively. In the $bb \bar{b} \bar q$ tetraquarks, the mass values lie around 15 GeV, specifically from 15564 to 15618 MeV for $bb \bar{b} \bar n$ tetraquarks and from 15669 to 15715 MeV for $bb \bar{b} \bar s$ tetraquarks. The mass splittings are 54 and 46 MeV for $bb \bar{b} \bar n$ and $bb \bar{b} \bar s$ tetraquarks, respectively. All the predicted masses for the $bb \bar{c} \bar q$ and $bb \bar{b} \bar{q}$ tetraquarks are above than their corresponding $S$-wave meson-meson thresholds. They can easily decay into two mesons via the fall-apart mechanism.

\begin{table}[h!]
\begin{center}
\caption{\label{Table:bbcbq}Predicted mass spectra for the $bb \bar c \bar q$ and $bb \bar b \bar q$ systems, corresponding thresholds and $\Delta$ values.  The meson masses are taken from \cite{ParticleDataGroup:2020ssz} and the predicted mass for $M_{B_c^\ast}=6338$ MeV is taken from \cite{Godfrey:1985xj}. All results are in units of MeV.}
\begin{tabular*}{18cm}{@{\extracolsep{\fill}}p{1cm}<{\centering}p{1.7cm}<{\centering}p{4.5cm}<{\centering}p{1.8cm}<{\centering}p{4.5cm}<{\centering}}
\hline\hline
 $J^{P}$  & Configuration                                             & Mass & $S$-wave threshold   & $\Delta$\\\hline

 $0^{+}$  & $|\{bb\}^{\bar 3}_1 (\bar c \bar n)^3_1 \rangle_0$    & \multirow{2}{*}{$\begin{bmatrix}11949 \\ 12088 \end{bmatrix}$}
               & \multirow{1}{*}{$\begin{bmatrix} \bar{B}_c \bar{B},\bar{B}_c^\ast \bar{B}^\ast   \end{bmatrix}$}  & \multirow{2}{*}{$\begin{bmatrix}(395, 286)\\(534, 425) \end{bmatrix}$}\\
            &  $|\{bb\}^6_0 (\bar c \bar n)^{\bar 6}_0\rangle_0$    \\
 $1^{+}$  &  $ |\{bb\}^{\bar 3}_1 (\bar c \bar n)^3_0\rangle_1  $    & \multirow{3}{*}{$\begin{bmatrix}11954\\11970 \\ 11981 \end{bmatrix}$}
               & \multirow{1}{*}{$\begin{bmatrix}  \bar{B}_c \bar{B}^\ast, \bar{B}_c^\ast \bar{B}^\ast  \end{bmatrix}$} &\multirow{3}{*}{$\begin{bmatrix}(355,291) \\(371,307) \\ (382,318) \end{bmatrix}$} \\ 
                 &  $|\{bb\}^{\bar 3}_1 (\bar c \bar n)^3_1\rangle_1$    \\
            &  $|\{bb\}^{6}_0 (\bar c \bar n)^{\bar 6}_1\rangle_1$    \\
 $2^{+}$  &   $|\{bb\}^{\bar 3}_1 (\bar c \bar n)^3_1\rangle_2$    & \multirow{1}{*} {$\begin{bmatrix}11987 \end{bmatrix}$}  &  \multirow{1}{*} {$\begin{bmatrix}\bar{B}_c^\ast \bar{B}^\ast \end{bmatrix}$}  &  \multirow{1}{*} {$\begin{bmatrix}324 \end{bmatrix}$} \\\hline

 $0^{+}$  & $|\{bb\}^{\bar 3}_1 (\bar c \bar s)^3_1 \rangle_0$    & \multirow{2}{*}{$\begin{bmatrix}12022 \\ 12143 \end{bmatrix}$}
               & \multirow{1}{*}{$\begin{bmatrix} \bar{B}_c \bar{B}_s,\bar{B}_c^\ast \bar{B}^\ast_s  \end{bmatrix}$} & \multirow{2}{*}{$\begin{bmatrix}(381, 269)\\(502, 390) \end{bmatrix}$}\\
            &  $|\{bb\}^6_0 (\bar c \bar s)^{\bar 6}_0\rangle_0$    \\
 $1^{+}$  &  $ |\{bb\}^{\bar 3}_1 (\bar c \bar s)^3_0\rangle_1  $    & \multirow{3}{*}{$\begin{bmatrix}12031\\12049\\ 12058 \end{bmatrix}$}
               & \multirow{1}{*}{$\begin{bmatrix} \bar{B}_c \bar{B}_s^\ast, \bar{B}_c^\ast \bar{B}_s  \end{bmatrix}$}  & \multirow{3}{*}{$\begin{bmatrix}(341, 278)\\(369,296) \\   (368, 305) \end{bmatrix}$}\\
                 &  $|\{bb\}^{\bar 3}_1 (\bar c \bar s)^3_1\rangle_1$    \\
            &  $|\{bb\}^{6}_0 (\bar c \bar s)^{\bar 6}_1\rangle_1$    \\
 $2^{+}$  &   $|\{bb\}^{\bar 3}_1 (\bar c \bar s)^3_1\rangle_2$    & \multirow{1}{*} {$\begin{bmatrix}12063 \end{bmatrix}$}  &  \multirow{1}{*} {$\begin{bmatrix} \bar{B}_c^\ast \bar{B}_s \end{bmatrix}$}  &  \multirow{1}{*} {$\begin{bmatrix}310 \end{bmatrix}$} \\ \hline \hline

 $0^{+}$  & $|\{bb\}^{\bar 3}_1 (\bar b \bar n)^3_1 \rangle_0$    & \multirow{2}{*}{$\begin{bmatrix}15564 \\ 15618 \end{bmatrix}$}
               & \multirow{1}{*}{$\begin{bmatrix} \eta_b \bar{B}, \Upsilon \bar{B}^\ast \end{bmatrix}$}  & \multirow{2}{*}{$\begin{bmatrix}(886, 779)\\(940, 833) \end{bmatrix}$}\\
            &  $|\{bb\}^6_0 (\bar b \bar n)^{\bar 6}_0\rangle_0$    \\
 $1^{+}$  &  $ |\{bb\}^{\bar 3}_1 (\bar b \bar n)^3_0\rangle_1  $    & \multirow{3}{*}{$\begin{bmatrix}15573 \\15588 \\ 15595  \end{bmatrix}$}
               & \multirow{1}{*}{$\begin{bmatrix} \eta_b \bar{B}^\ast, \Upsilon \bar{B}^\ast  \end{bmatrix}$}  & \multirow{3}{*}{$\begin{bmatrix}(850, 788)\\(865, 803) \\ (872, 810) \end{bmatrix}$} \\
                 &  $|\{bb\}^{\bar 3}_1 (\bar b \bar n)^3_1\rangle_1$    \\
            &  $|\{bb\}^{6}_0 (\bar b \bar n)^{\bar 6}_1\rangle_1$    \\
 $2^{+}$  &   $|\{bb\}^{\bar 3}_1 (\bar b \bar n)^3_1\rangle_2$    &  \multirow{1}{*} {$\begin{bmatrix}15607 \end{bmatrix}$}  &   \multirow{1}{*} {$\begin{bmatrix}\Upsilon \bar{B}^\ast \end{bmatrix}$}  &  \multirow{1}{*} {$\begin{bmatrix} 822\end{bmatrix}$}  \\ \hline

 $0^{+}$  & $|\{bb\}^{\bar 3}_1 (\bar b \bar s)^3_1 \rangle_0$    & \multirow{2}{*}{$\begin{bmatrix}15669\\ 15715 \end{bmatrix}$}
               & \multirow{1}{*}{$\begin{bmatrix}\eta_b \bar{B}_s, \Upsilon \bar{B}_s^\ast  \end{bmatrix}$}  & \multirow{2}{*}{$\begin{bmatrix}(903, 793)\\(949, 839) \end{bmatrix}$}\\
            &  $|\{bb\}^6_0 (\bar b \bar s)^{\bar 6}_0\rangle_0$    \\
 $1^{+}$  &  $ |\{bb\}^{\bar 3}_1 (\bar b \bar s)^3_0\rangle_1  $    & \multirow{3}{*}{$\begin{bmatrix}15672\\15683\\ 15685  \end{bmatrix}$}
               & \multirow{1}{*}{$\begin{bmatrix} \eta_b \bar{B}_s^\ast, \Upsilon \bar{B}_s^\ast \end{bmatrix}$}  & \multirow{3}{*}{$\begin{bmatrix}(858, 796)\\(869, 807) \\ (871, 809) \end{bmatrix}$}\\
                 &  $|\{bb\}^{\bar 3}_1 (\bar b \bar s)^3_1\rangle_1$    \\
            &  $|\{bb\}^{6}_0 (\bar b \bar s)^{\bar 6}_1\rangle_1$    \\
 $2^{+}$  &   $|\{bb\}^{\bar 3}_1 (\bar b \bar s)^3_1\rangle_2$    & \multirow{1}{*} {$\begin{bmatrix} 15691 \end{bmatrix}$}  &  \multirow{1}{*} {$\begin{bmatrix} \Upsilon B_s^\ast \end{bmatrix}$}  &  \multirow{1}{*} {$\begin{bmatrix} 815 \end{bmatrix}$} \\

\hline\hline
\end{tabular*}
\end{center}
\end{table}

We compare our mass results with the predictions of other studies in Table \ref{Table:6}. For the $bb\bar{c}\bar{n}$ tetraquarks, we see that our results agree with those of Refs. \cite{Lu:2021kut,Chen:2016ont}. Our predicted masses are quite higher than masses of Ref. \cite{Weng:2021ngd}. For the $bb\bar{c}\bar{s}$ tetraquarks, our results agree well with mass values of Ref. \cite{Lu:2021kut}. Mass predictions of Ref. \cite{Chen:2016ont} are quite higher whereas predictions of  Ref. \cite{Weng:2021ngd} are quite lower than our results. In the case of $bb\bar{b}\bar{n}$ states, our results agree with results of Ref. \cite{Chen:2016ont}. Mass predictions of Ref. \cite{Jiang:2017tdc,Lu:2021kut,Weng:2021ngd} are rather different than our predicted mass values. This argument can be transferred to the $bb\bar{b}\bar{s}$ tetraquark states.

\begin{table}[h!]
\caption{\label{Table:6} Comparison of $bb\bar{c}\bar{q}$ and $bb\bar{b}\bar{q}$ tetraquark masses with other studies.  All results are in units of MeV except Ref. \cite{Jiang:2017tdc} where they are in GeV. }
\begin{ruledtabular}
\begin{tabular}{cccccccccc}
Configuration& $J^P$&This work&\cite{Chen:2016ont}&\cite{Jiang:2017tdc} &\cite{Lu:2021kut}&\cite{Weng:2021ngd} \\
\hline
 $bb\bar{c}\bar{n}$&  &  &  &  \\
 \hline
 &$0^+$&11949-12083&12144.8-12257.2& $\cdots$ & 11906-12013 & 11581.7-11694.9\\

 &$1^+$&11954-11970-11981& 12147.6-12196.5-12219.1& $\cdots$ & 11873-11933-11984 & 11582.2-11624.7-11658.6\\

 &$2^+$&11987&12225.2 & $\cdots$ & 11948 & 11651.7\\ \hline

$bb\bar{c}\bar{s}$&  &  &  &  \\ \hline

 &$0^+$&12022-12143&12322.7-12439.4& $\cdots$ & 11985-12067 & 11675.1-11787.2\\

 &$1^+$&12031-12049-12058& 12322.1-12375.8-12396.5& $\cdots$ & 11950-12003-12040 & 11673.3-11722.2-11746.3\\

 &$2^+$&12063& 12405.9 & $\cdots$ & 12018 & 11750.2\\ \hline \hline

 $bb\bar{b}\bar{n}$&  &  &  &  \\ \hline

 &$0^+$&15564-15618&15468.6-15572.7 & $13.5\pm0.4-13.7\pm0.3$ & 15109-15158 & 14706.1-14850.9\\

 &$1^+$&15573-15588-15595& 15493.5-15519.7-1553.4& $13.5\pm0.4-13.5\pm0.4$ & 15107-15137-15142& 14712.1-14485.1-14851.6\\

 &$2^+$&15607&15545.0& $\cdots$ & 15154 & 14871.8\\ \hline 
 
 $bb\bar{b}\bar{s}$&  &  &  &  \\ \hline

 &$0^+$&15669-15715& 15752.5-15645.3& $\cdots$ & 15185-15232 & 14793.1-14936.2\\

 &$1^+$&15672-15683-15685& 15672.2-15697.0-15733.3& $\cdots$ & 15184-15213-15219& 14804.5-14929.2-14943.7\\

 &$2^+$&15691& 15723.9& $\cdots$ & 15231 & 14956.6\\ 

\end{tabular}
\end{ruledtabular}
\end{table}

Considering color configurations, QCD sum rule analysis of Ref.  \cite{Jiang:2017tdc} gave mass values with $J^P=0^+$ as $13.5 \pm 0.4$ GeV for $6 \otimes \bar{6}$ type interpolating current and $13.7\pm 0.3$ GeV for $ \bar{3} \otimes 3$ type interpolating current of $bb\bar{b}\bar{q}$ tetraquarks. For the $J^P=1^+$, both $6 \otimes \bar{6}$ and $\bar{3} \otimes 3$ type interpolating currents gave mass value as $13.5 \pm 0.4$ GeV for the $bb\bar{b}\bar{q}$ tetraquarks. In Ref. \cite{Weng:2021ngd}, the ground states of $nb \bar b \bar b$ and $s b \bar b \bar b$ have $J^P=0^+$ quantum number and are dominated by $6 \otimes \bar{6}$ color configurations with a rate of $(>85\%)$. In the $nc \bar b \bar b$ and $sc \bar b \bar b$ states, $\bar{3} \otimes  3$ color configurations dominate  $J^P=0^+$ ground states with a rate of $(>70\%)$. In Ref.  \cite{Lu:2021kut}, ground states of $bb\bar{c}\bar{q}$ and $bb\bar{b}\bar{q}$ tetraquarks are in the $\bar{3} \otimes  3$ configurations with $J^P=1^+$ whereas in our results ground states of $bb\bar{c}\bar{q}$ and $bb\bar{b}\bar{q}$ tetraquarks are in 
$6 \otimes \bar{6}$  color configurations but with $J^P=0^+$.  

Except some states in the $cc\bar c \bar q$ sector, all the states of $cc \bar b \bar q$, $bb \bar c \bar q$ and $bb \bar b \bar q$ sectors are above than their corresponding two-meson thresholds. They may decay into two mesons through the fall-apart mechanism. Although these states are not stable, some of them may be observed in the future experiments as resonances with finite decay widths. In addition to fall-apart decay mechanism, triply-heavy tetraquark states may decay through other two-body strong decay modes. 
In the hidden charm and hidden bottom states, $c\bar c$ and $b \bar b$ quark pairs may decay into the meson-meson or even baryon-antibaryon final states.  

\begin{figure}[H]
\begin{center}
 [$cc \bar c \bar n$ ]{\includegraphics[totalheight=8cm,width=8cm]{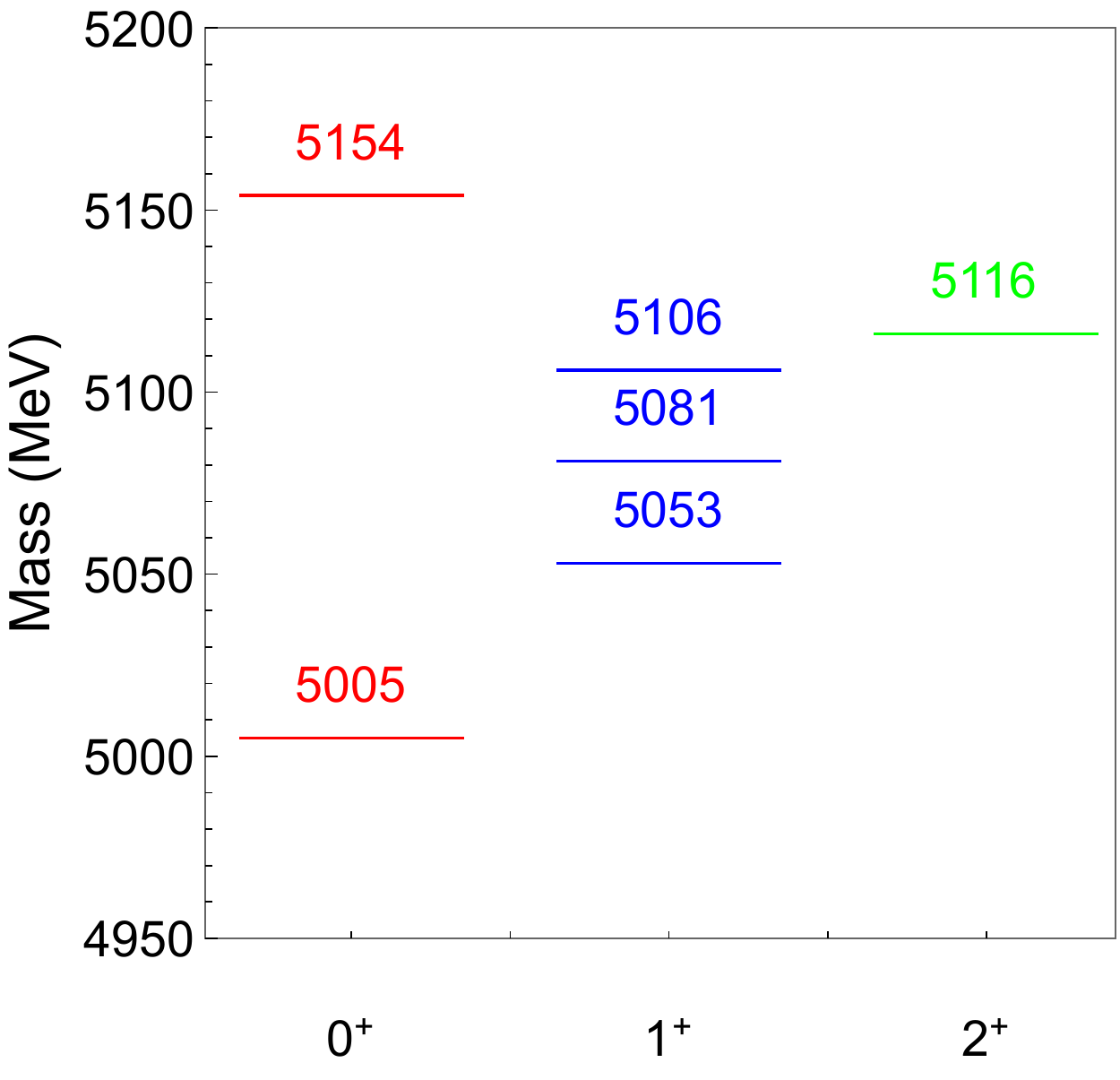}}
[$cc \bar c \bar s$ ]{\includegraphics[totalheight=8cm,width=8cm]{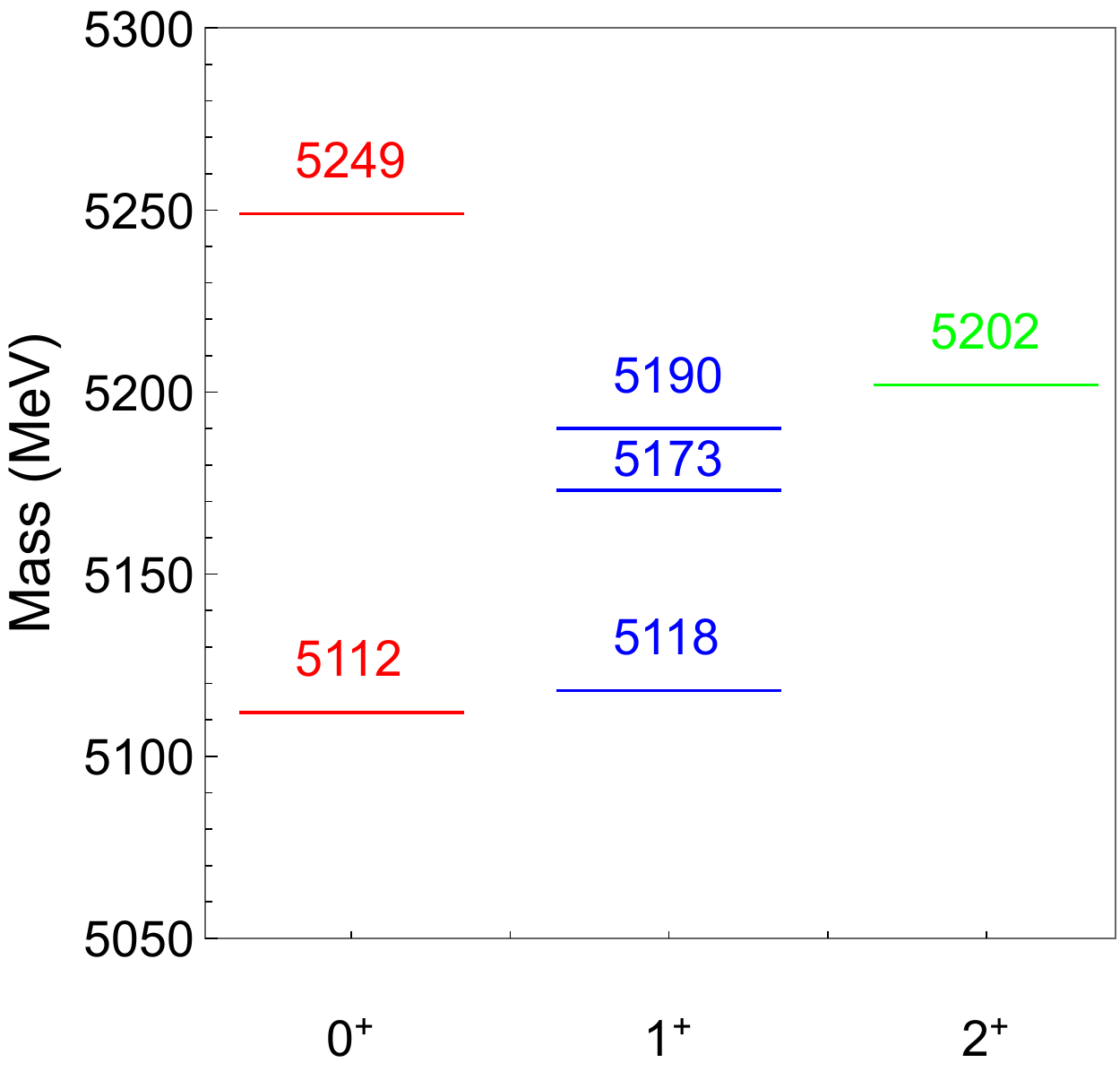}}
[$cc \bar b \bar n$ ]{\includegraphics[totalheight=8cm,width=8cm]{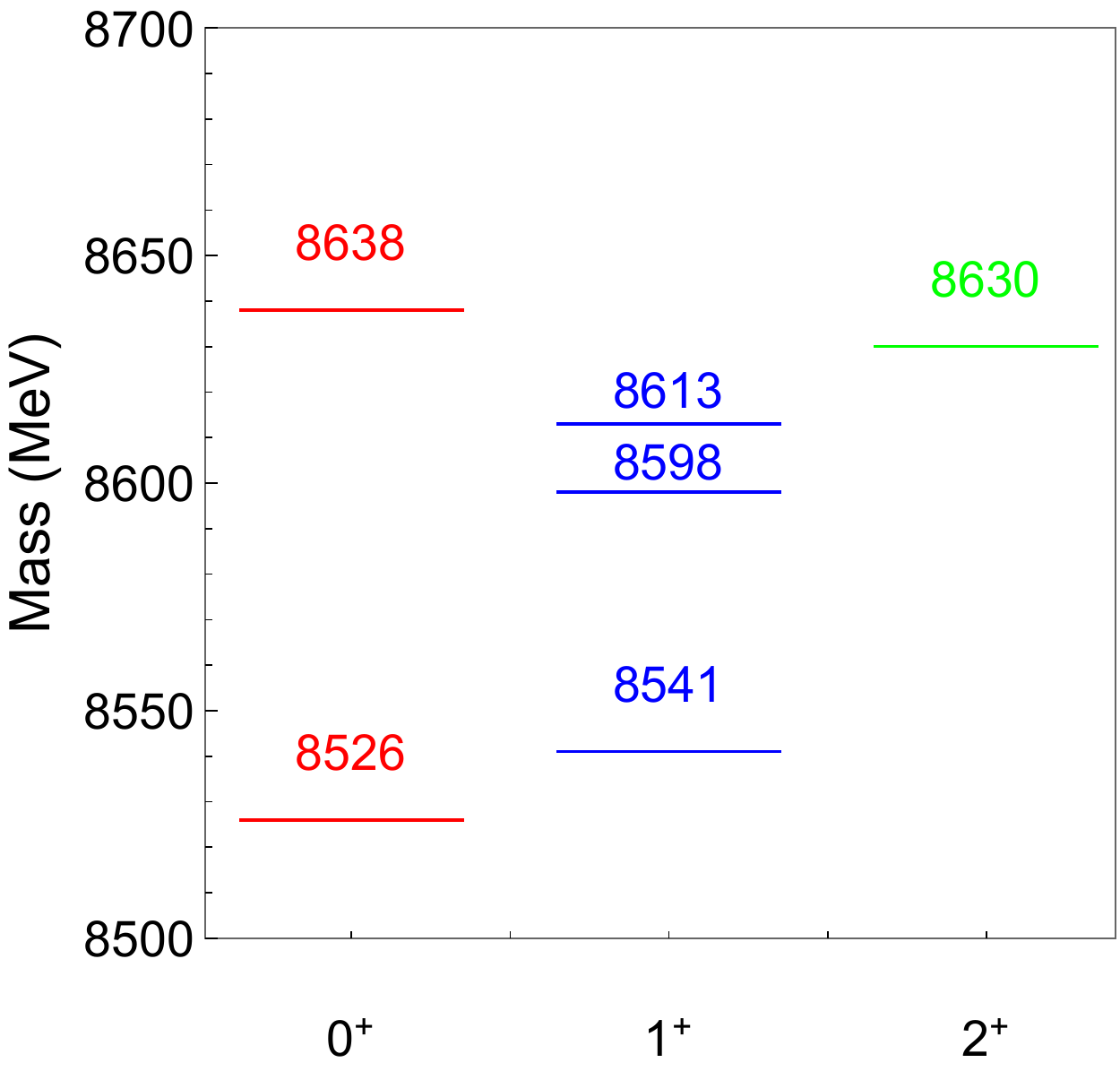}}
[$cc \bar b \bar s$ ]{\includegraphics[totalheight=8cm,width=8cm]{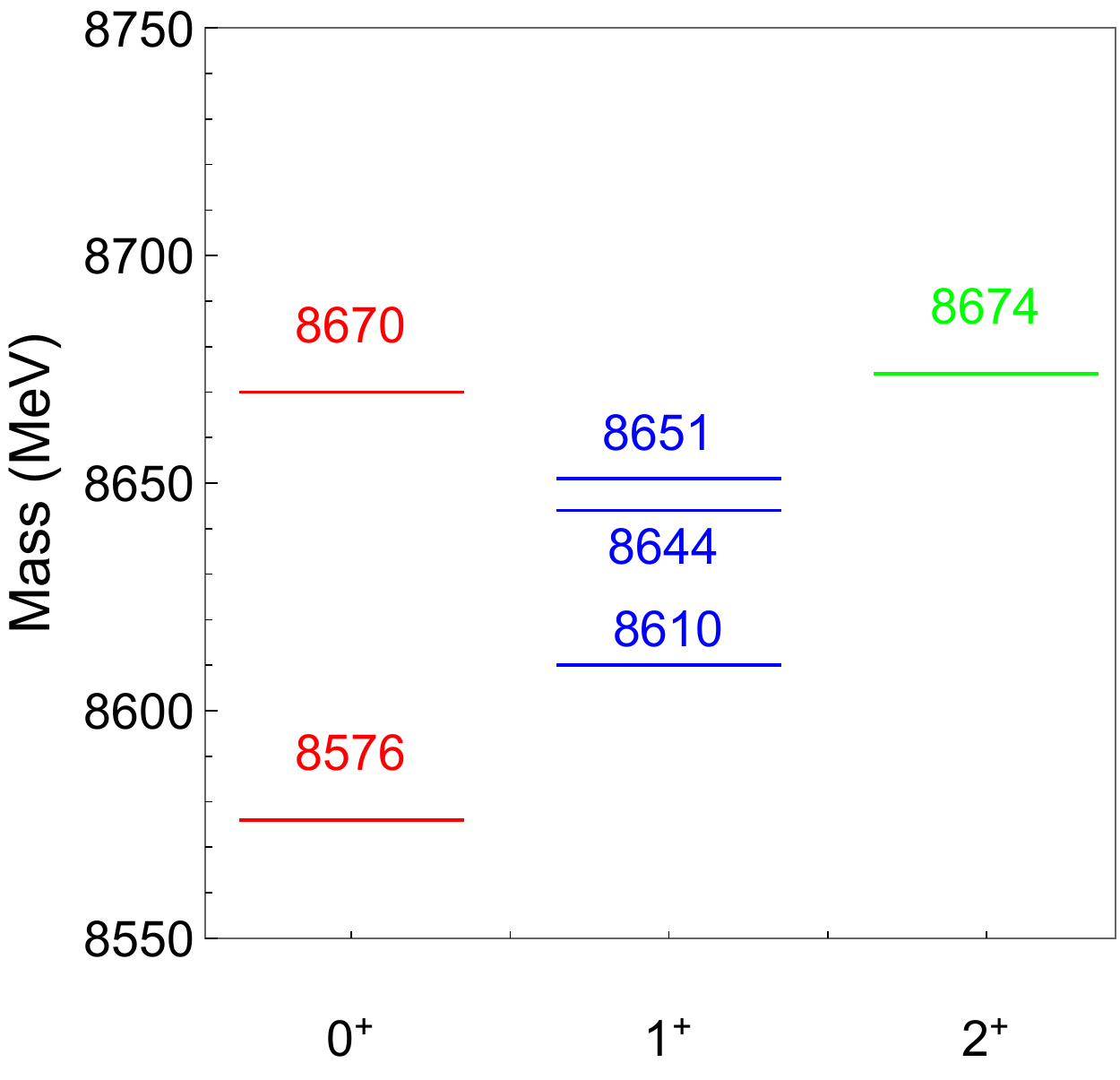}}
\end{center}
\caption{Predicted mass spectra of $cc \bar c \bar q$ and $cc \bar b \bar q$ states}
\label{fig:Mass1}
\end{figure}

\begin{figure}[H]
\begin{center}
[$bb \bar c \bar n$ ]{\includegraphics[totalheight=8cm,width=8cm]{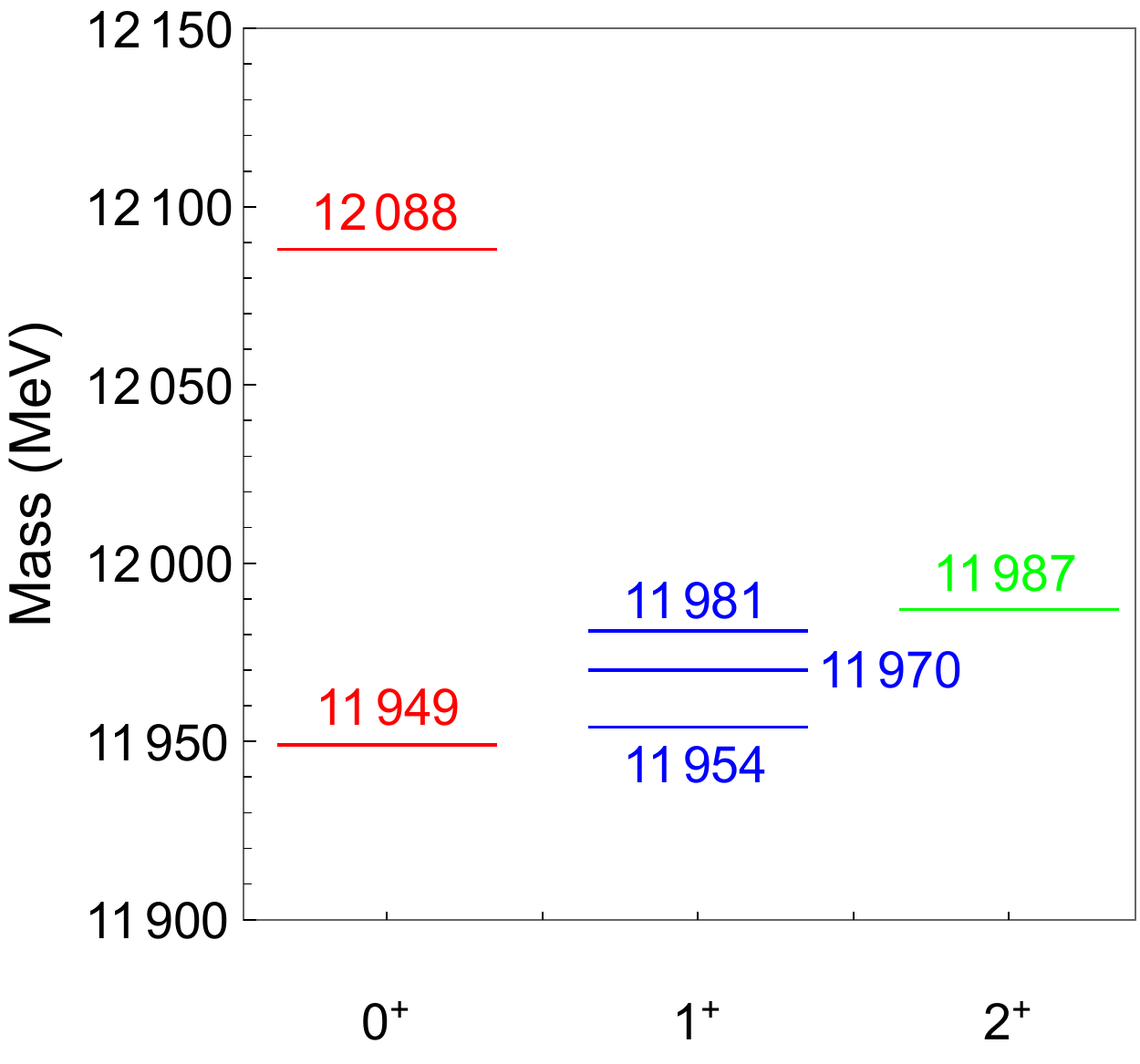}}
[$bb \bar c \bar s$ ]{\includegraphics[totalheight=8cm,width=8cm]{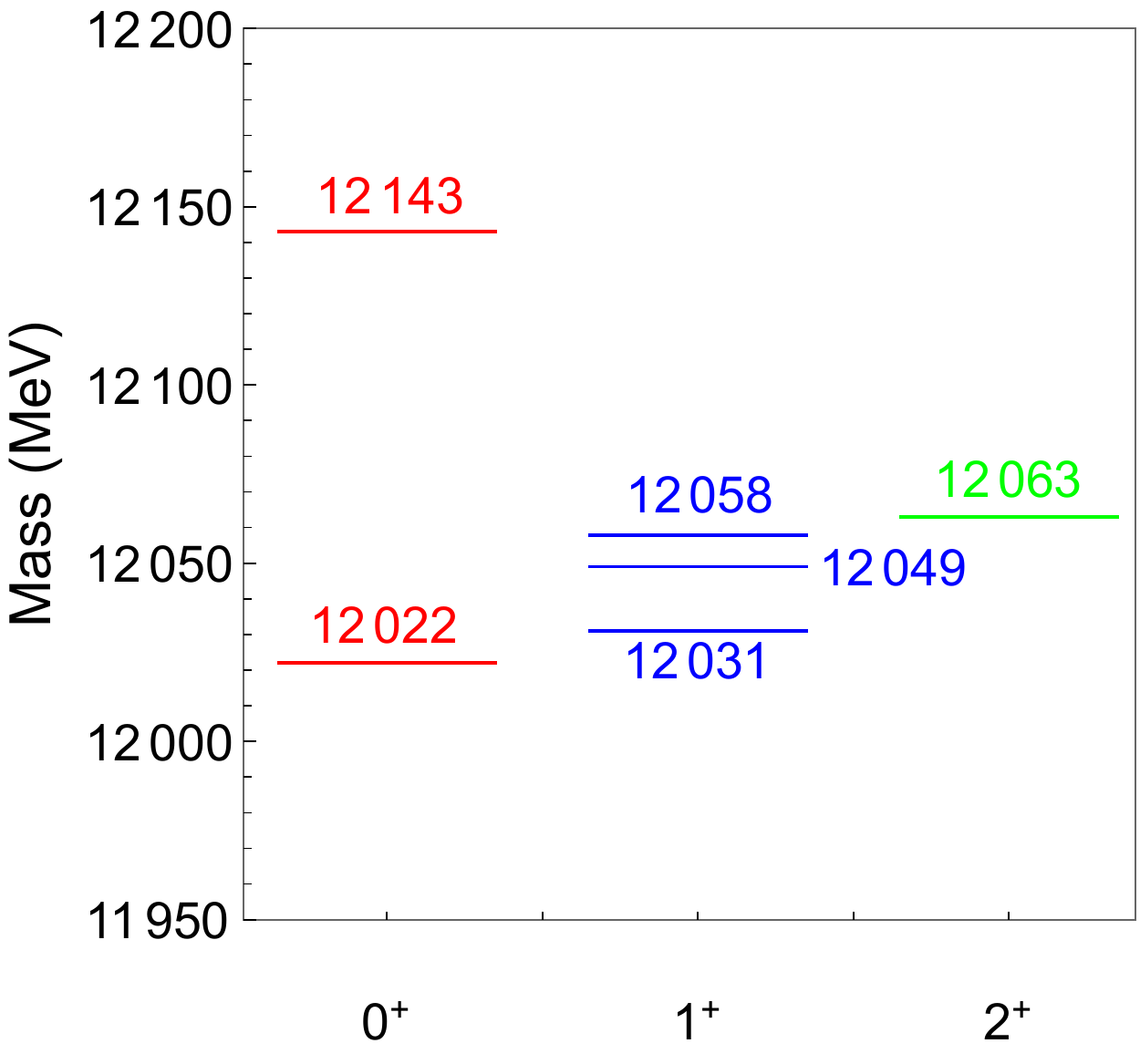}}
[$bb \bar b \bar n$ ]{\includegraphics[totalheight=8cm,width=8cm]{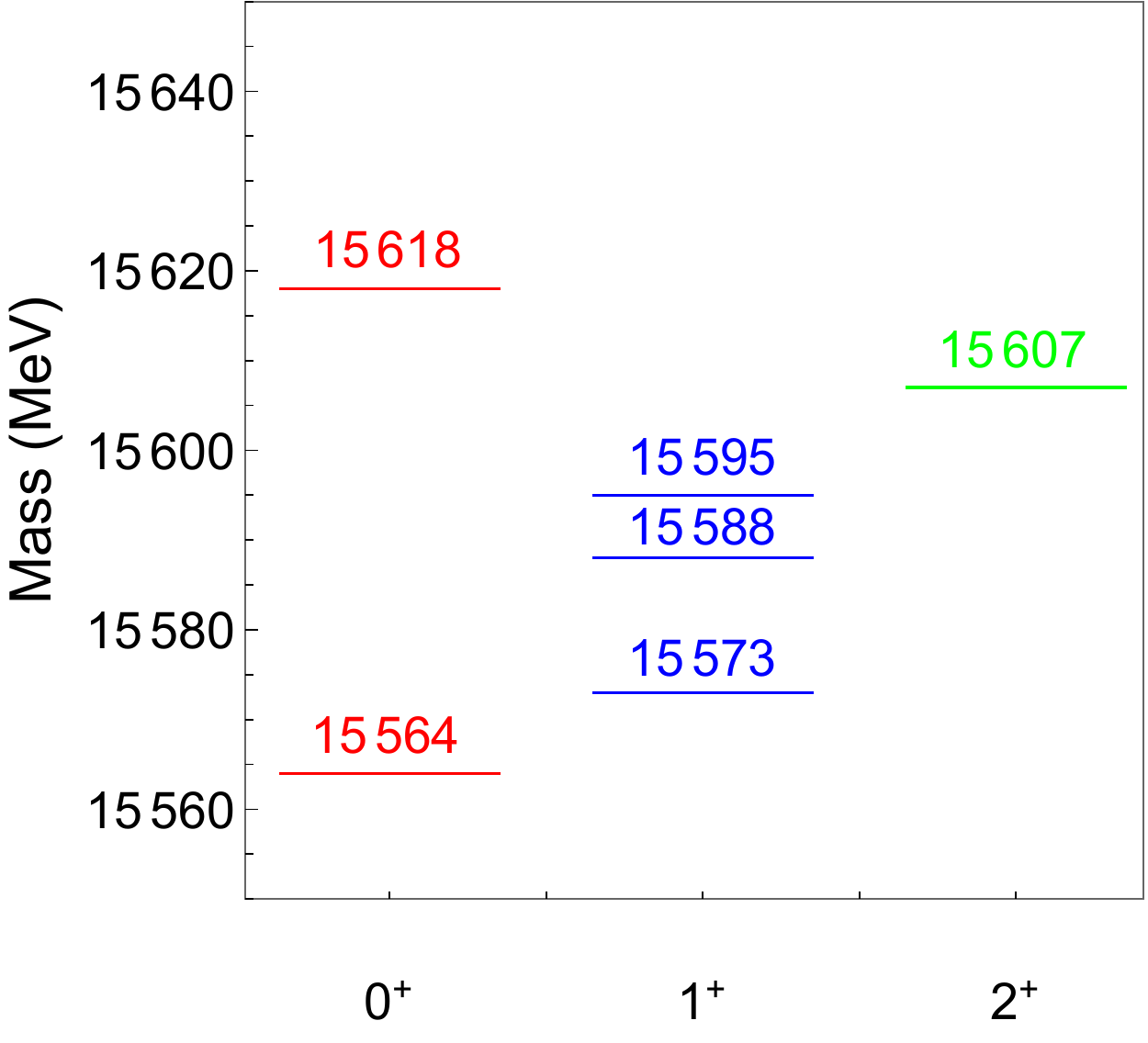}}
[$bb \bar b \bar s$ ]{\includegraphics[totalheight=8cm,width=8cm]{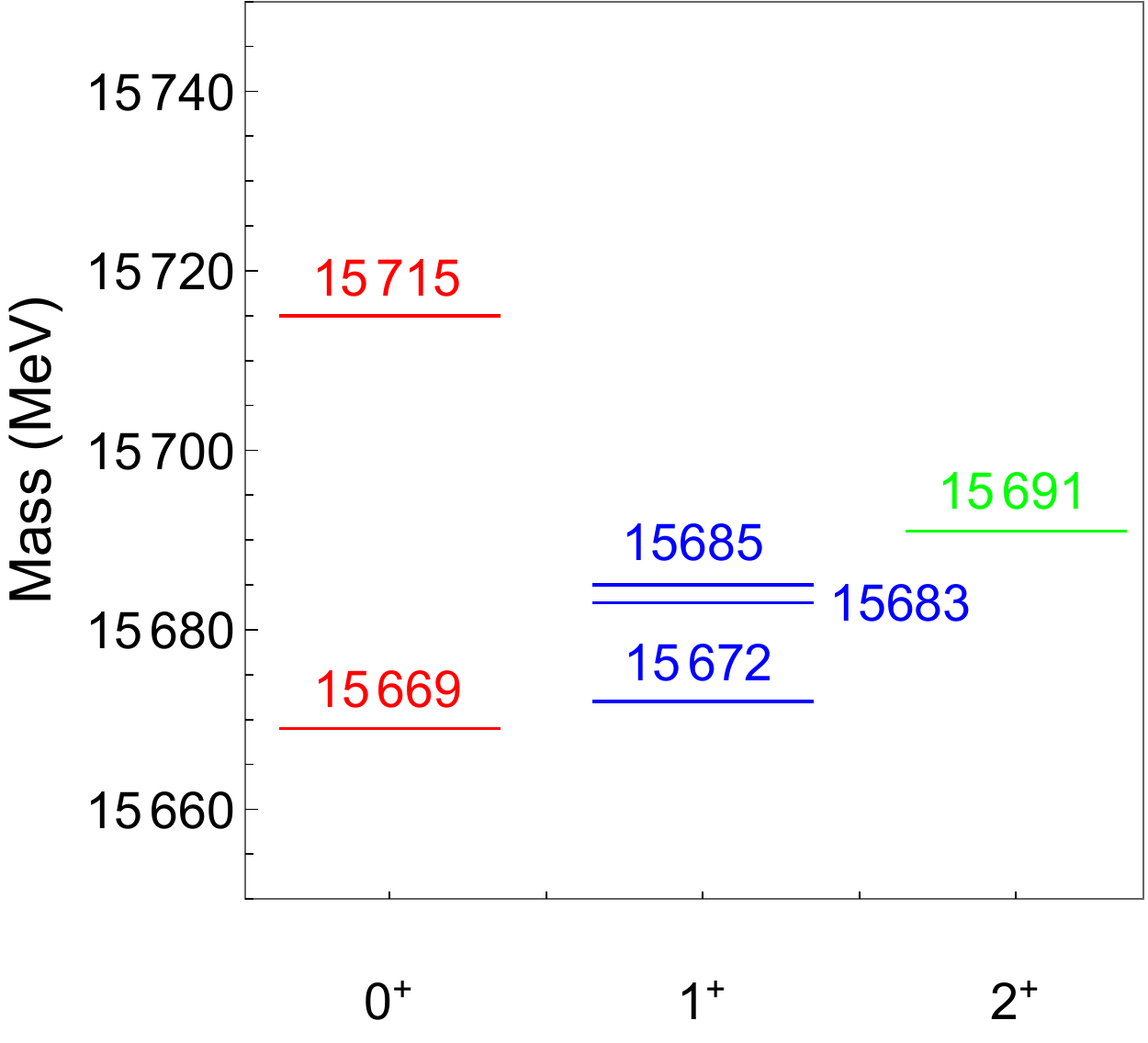}}
\end{center}
\caption{Predicted mass spectra of $bb \bar c \bar q$ and $bb \bar b \bar q$ states}
\label{fig:Mass2}
\end{figure}

Before ending this section, mass spectra of $S$-wave triply-heavy tetraquark states are given in Figs. \ref{fig:Mass1} and \ref{fig:Mass2}. We can see from both figures that they have similar mass patterns. This is an indication of light flavor $SU(3)$ symmetry and heavy quark symmetry which have been successfully used in traditional hadrons as well as in singly-heavy tetraquarks and doubly-heavy tetraquarks. This heavy quark symmetry can be an ideal tool for investigating triply-heavy tetraquarks. In the heavy quark symmetry if one increase the mass of heavy quark, technically in the $m_Q \to \infty$ limit, QCD can be handled independent of the heavy quark flavor and spin. The recently observed $T_{cc}^ +$ state has a quark content of $cc \bar u \bar d$ and is the longest living exotic hadron discovered till now \cite{LHCb:2021vvq,LHCb:2021auc}. For the doubly-heavy tetraquarks $Q_1Q_2 \bar q_3 \bar q_4$, it was shown in Refs. \cite{Vijande:2009kj,Hernandez:2019eox,Lu:2020rog} that the mass ratio between the heavy $Q_1Q_2$ diquark and $\bar q_3 \bar q_4$ light antidiquark is one of the determining factors for being a stable tetraquark. For example, if we replace charm quarks with bottom quarks in $T_{cc}^ +$ state, the forming $bb \bar u \bar d$ state has the largest mass ratio between heavy quarks and light antiquarks and is a compact stable tetraquark candidate suggested by Ref. \cite{Karliner:2017qjm} after the discovery of $\Xi_{cc}^{++}$ baryon \cite{LHCb:2017iph}. If we replace $\bar u$ or $\bar d$ quark with $\bar c$ or $\bar b$ in the $bb \bar u \bar d$ state, the ratio between heavy quarks and antiquarks decrease. This may be another indication that why $bb \bar c \bar q$ and $bb \bar b \bar q$ triply-heavy tetraquark states cannot be stable. Contrary to this, some bound state candidates in the $bb \bar c \bar q$ sector was found in Ref. \cite{Liu:2022jdl}. In the same work, a bound state candidate in the $cc \bar c \bar q$ was found which supports our result in the $cc \bar c \bar q$ sector.


\section{\label{sec:level5}Summary and Concluding Remarks}
In this work, we consider flavor exotic triply-heavy tetraquarks as composed of diquarks and antidiquarks and systematically study $S$-wave triply-heavy tetraquarks by solving Schrödinger equation numerically. The calculations are carried out in a nonrelativistic quark model with a color interaction described by a potential computed in AdS/QCD. The model used in this paper consists of a central potential  $V_{AdS}$ which contains a linear part at the large distances for confining and the short-distance behaviour of perturbative QCD, spin dependent term $V_{spin}$, and a constant term $V_0$.  

From the predicted mass spectra, we can see that in the $cc\bar{c}\bar{q}$ systems we have found narrow state candidates. For $cc\bar{b}\bar{q}$, $bb\bar{c}\bar{q}$, and $bb\bar{b}\bar{q}$  systems all the triply-heavy tetraquarks lie above the corresponding meson-meson thresholds, therefore no stable state exists in this system. The mass spectra for the flavor exotic triply-heavy tetraquark states have quite similar patterns preserving the light flavor $SU(3)$ and heavy quark symmetries. 
Our results are quite different than Refs. \cite{Chen:2016ont,Jiang:2017tdc,Lu:2021kut,Weng:2021ngd} since we have some bound state candidates in the $cc \bar c \bar q$ tetraquark configurations. These studies are conducted with different models explained before. The differences among these studies stem from the different choices of interactions, parameters or assumptions in the models. It should be also mentioned that up to now $B_c^\ast$ meson has not been observed in the experiments. Therefore a precise measurement of its mass is important for the corresponding threshold values. 

For the $XYZ$ states, it is not clear that whether a tetraquark or molecule configuration is plausible for explaining even the same particles, especially one or more light quarks do have place in these states. In this manner triply-heavy tetraquark states may open a way to identify genuine tetraquark states. Furthermore, triply-heavy tetraquarks are intermediate states between the doubly- and fully-heavy tetraquarks due to the mass relations of quarks, $m_{q}{\ll}m_{c}{\ll}m_{b}$. The binding force of the triply-heavy tetraquark states is dominantly provided by the gluon exchange mechanism. Besides the phenomenological models relying on gluon exchange interactions, other nonperturbative methods such as lattice QCD, effective field theories, and QCD sum rule can be used for further studies. For example, lattice QCD calculation is relatively easy for the $bb \bar c \bar q$ abd $cc \bar b \bar q$ triply-heavy tetraquarks since annihilation effects of the same flavor quark does not contaminate the simulation. 

The studies regarding with these triply-heavy tetraquark states may help to understand underlying structures of the flavor exotic and multiquark states. We hope that our results can be of help for the future experimental studies.

\begin{acknowledgments}
The author of this work would like to thank Soner Albayrak, Rodrigo C. L. Bruni and Henrique Boschi-Filho for their valuable discussions about AdS/QCD potential. The author also thanks to the anonymous referee for his/her useful comments.
\end{acknowledgments}

\bibliography{triply-heavy-tetraquark}

\end{document}